\documentclass[epj]{svjour}
\usepackage{graphicx}
\usepackage{amssymb}
\usepackage{dcolumn}
\usepackage{amsmath}
\usepackage{bm}
\usepackage{textcomp}
\usepackage{epstopdf}
\usepackage{color,url}
\usepackage{ulem}
\usepackage{soul}
\usepackage[toc,page,title,titletoc,header]{appendix}
\usepackage[colorlinks,
            linkcolor=blue,
            anchorcolor=blue,
            citecolor=blue,
            urlcolor=blue
            ]{hyperref}
\usepackage{txfonts}
\usepackage{caption}

\begin{document}

\title{Non-equilibrium cumulants within model A from crossover to first-order phase transition side}
\author{Lijia Jiang\inst{1} \and
Jingyi Chao\inst{2}}
\institute{Institute of Modern Physics, Northwest University, 710127 Xi'an, China \and
College of Physics and Communication Electronics, Jiangxi Normal University, Nanchang 330022, China}

\date{Received: date / Revised version: date}

\abstract{We study the non-equilibrium cumulants of the chiral order parameter field ($\sigma$ field) in different phase transition scenarios via Langevin dynamics. Cumulants up to fourth-order have been calculated based on the spacetime-dependent $\sigma$ configurations from the event-by-event numerical simulations. By limiting the cooling of the system in a Hubble-like way, the out-of-equilibrium cumulants illustrate clear memory effects during the evolution. Both the signs and the magnitudes of the high-order cumulants differ from the equilibrium ones below the phase transition temperature. Especially, the dynamical cumulants grow more intensively from the first-order phase transition side than they do from the crossover side. In addition, analysis of the high-order off-equilibrium cumulants on the hypothetical freeze-out lines present non-monotonic curves in the large chemical potential region.
}

\maketitle

\section{Introduction}
Searching for the location of the critical point and depicting the QCD phase diagram is one of the fundamental topics in heavy-ion physics, which has been extensively studied for decades~\cite{Klevansky:1992qe,Roberts:1994dr,Berges:2000ew,Karsch:2003jg,Fukushima:2003fw,Aoki:2006we,Fu:2007xc,Aoki:2009sc,STAR:2010vob,Qin:2010nq,Jiang:2013yoa,Luo:2017faz}. Induced by the divergent fluctuations at the critical point, dramatic increases of the cumulants of the final state proton multiplicities in the critical region are predicted as a consequence of strong coupling between the quark matter and the chiral order parameter field (the $\sigma$ field)~\cite{Stephanov:1998dy,Hatta:2003wn,Stephanov:2008qz,Athanasiou:2010kw,Kitazawa:2012at}. Even though the correlation length of the chiral order parameter will be suppressed by the finite size of the QCD fireball in heavy-ion collisions, the high-order correlators remain quite sensitive to the increase of the correlation length near the critical point since they are proportional to the high power of the correlation length~\cite{Stephanov:2008qz}. Especially, the quartic critical cumulant is predicted to be negative as approaching the critical point from the crossover side, but positive from the first-order phase transition side, thus the non-monotonic kurtosis along the chemical freeze-out line is expected to carry the signals of the QCD critical point in laboratory observations~\cite{Stephanov:2008qz,Asakawa:2009aj,Friman:2011pf,Stephanov:2011pb}.

The experimental exploration of the QCD phase diagram is performed at the BNL Relativistic Heavy Ion Collider (RHIC). The STAR collaboration has measured the high-order cumulants of net proton production in Au+Au collisions for the collision energy ranging from $7.7$ GeV to $200$ GeV~\cite{STAR:2010vob,STAR:2013gus,Luo:2015ewa}. The experimental data of $\kappa \sigma ^{2}$ ($=C_{4}/C_{2}$) shows a large deviation from the baselines determined by Poisson statistics and presents complex non-monotonic structure in the collision energy region below $39$ GeV ~\cite{Luo:2015ewa,STAR:2020tga}, which has not  been fully explained in theoretical calculations until now. Meanwhile, more experimental measurements are being executed by the HADES collaboration at GSI~\cite{HADES:2020wpc} and by the ALICE Collaboration at the LHC~\cite{ALICE:2019nbs}. Future experimental facilities such as FAIR in Darmstadt, NICA in Dubna, HIAF in Huizhou and J-PARC in Tokai will further explore the details of the QCD phase transition at the (relative) low collision energy region.

Since the high-order cumulants of the protons are measured after hadrons chemically freeze-out, to identify the signals of the critical point from the experimental data, a proper freeze-out scheme near the critical point is introduced to the hydrodynamic model and the equilibrium cumulants of protons on the freeze-out surface are investigated ~\cite{Jiang:2015hri,Jiang:2015cnt,Jiang:2016duk}.
The resulted cumulants qualitatively describe the acceptance dependence of the experimental data and roughly fits the $C_{4}$ and $\kappa \sigma ^{2}$ data, but  $C_{2}$ and $C_{3}$ are overestimated in this framework. Incontrovertibly, the  off-equilibrium dynamics play an essential role in the dynamical evolution with phase transition to address the critical slowing down near a critical point~\cite{Wu:2021xgu,Berdnikov:1999ph} and to reveal the domain formation at the first-order phase transition~\cite{Randrup:2010ax}. Indeed, the high-order cumulants vary significantly, not only in the magnitudes but also the signs, in the real-time evolution compared with the equilibrium hypothesis~\cite{Mukherjee:2015swa,Jiang:2017sni,Jiang:2017fas,Jiang:2017mji}. Furthermore, the early time fluctuations and the critical enhancements around the critical point can be probed by the rapidity-window-dependent Gaussian cumulant ~\cite{Sakaida:2017rtj}. In Ref.~\cite{Nahrgang:2018afz}, it shows that the effects of the nonlinear coupling and finite size are manifested through the reduction of the correlation length near the pseudo-critical temperature. In Ref.~\cite{Rougemont:2018ivt}, the well separated equilibrating time of non-hydrodynamic quasi-normal modes in different channels is investigated at the critical point by a phenomenological realistic holographic model, likely to be model B.

According to the classification of dynamic universality, the QCD chiral phase transition in heavy-ion collisions belongs to Model H, since it is governed by five conserved hydrodynamic modes~\cite{Rev1977,Ma1976,Son:2004iv,Chao:2020kcf,Nahrgang:2020yxm,De:2020yyx,Schweitzer:2021iqk}. Dynamical models with hydrodynamic background, such as chiral hydrodynamics~\cite{Paech:2003fe,Nahrgang:2011mg,Nahrgang:2011vn,Herold:2013bi,Herold:2014zoa} and Hydro+~\cite{Stephanov:2017ghc,Rajagopal:2019xwg,Du:2020bxp} have been developed to describe the medium expansion and the critical mode evolution of QCD phase transition in heavy-ion collisions. However, besides the numerical complexities of these models, the equation of state, the  transport coefficients, and the freeze-out scheme in the vicinity of the critical point still needs careful discussions. The numerical calculations of cumulants are even more challenging when the long range correlation of high-order fluctuations~\cite{An:2020vri} and the nonzero-momentum critical modes are considered for different phase transition scenarios, thus it is worth employing the simplified relaxational model A and/or diffusive model B ~\cite{Mukherjee:2015swa,Jiang:2017sni,Jiang:2017fas,Jiang:2017mji,Sakaida:2017rtj,Nahrgang:2018afz,Rougemont:2018ivt,Nahrgang:2020yxm} to study the dynamical phase transition at a broader chemical potential regions.

Based on the event-by-event Langevin equation, we detailed study the dynamical evolution of the $\sigma$ field's cumulants in different phase transition scenarios and on the hypothetical freeze-out line.
Some preliminary results of the dynamical cumulants in various phase transition scenarios are provided in the proceedings~\cite{Jiang:2017fas,Jiang:2017mji}. Here, more calculations and discussions are presented, including those involving the impact of the initial temperature on evolution and the dynamical cumulants on the hypothetical freeze-out lines.
The paper is organized as follows: In Sec.~\ref{sec:model}, we introduce the dynamical model in detail where the necessary input parameters are exhibited to solve the stochastic equation, including the initial profile, expansion routine of temperature, damping coefficient, and the magnitude of noise.
In Sec.~\ref{sec:result}, we present the numerical results of $\sigma$ field's cumulants along the given evolution trajectories in both the crossover and the first-order phase transition scenario.
Afterward, we illuminate the results of non-equilibrium cumulants on the hypothetical freeze-out lines. In Sec.~\ref{sec:con}, we close the article by summarizing our main results and discussing future developments.

\section{Model and set-ups}\label{sec:model}
Linear sigma model is a QCD-inspired low energy effective theory, depicting the phase structure of strongly interacting matter in the $\mu$-$T$ plane via the order parameter $\sigma = \left\langle\bar{\psi}\psi\right\rangle$~\cite{Jungnickel:1995fp,Skokov:2010sf}.
As the mass of the $\sigma$ field approaches to zero near the critical point, its correlation length grows to infinity. The corresponding equation of motion of the long wavelength mode is well described by the Langevin equation~\cite{Nahrgang:2011mg,Xu:1999aq,Meistrenko:2020nwx}:
\begin{equation}
\partial ^{\mu }\partial _{\mu }\sigma \left( t,x\right) +\eta \partial
_{t}\sigma \left( t,x\right) +\frac{\delta V_{\mathrm{eff}}\left( \sigma \right) }{%
\delta \sigma }=\xi \left( t,x\right),
\label{langevin}
\end{equation}%
where the effective potential of the $\sigma$ field is explicitly written as
\begin{equation}
V_{\mathrm{eff}}\left( \sigma \right) =U\left( \sigma \right) +\Omega _{\bar{q}%
q}\left( \sigma \right).
\label{effpot}
\end{equation}%
$U\left( \sigma \right) $ denotes the vacuum contribution and takes the form of
\begin{equation}
U\left( \sigma \right) =\frac{\lambda ^{2}}{4}\left( \sigma
^{2}-v^{2}\right) ^{2}-h_{q}\sigma -U_{0}.
\end{equation}
The parameters $\lambda$, $\sigma$, $h_q$, and $U_0$ are set by the hadrons properties at zero temperature. As the chiral symmetry is spontaneously broken in the vacuum, the nonzero expectation of the $\sigma$ field is $\left\langle \sigma \right\rangle =f_{\pi }=93$ MeV with $%
\left\langle \vec{\pi}\right\rangle =0$. In reality, the chiral
symmetry is explicitly broken by the light quark mass, then  the linear term is included with $%
h_{q}=f_{\pi }m_{\pi }^{2}$ and $m_{\pi }=138$ MeV.
$\nu ^{2}=f_{\pi}^{2}-m_{\pi }^{2}/\lambda ^{2}$ and the mass of  $\sigma$ is $m_{\sigma} \sim 600$ MeV by setting $\lambda ^{2}=20$. The zero-point energy $%
U_{0}=m_{\pi }^{4}/\left( 4\lambda ^{2}\right) -f_{\pi }^{2}m_{\pi }^{2}$.
Note that we have neglected the meson fluctuations of $\vec{\pi}$, since the mass of the triplet is finite in the critical regime.
$\Omega _{\bar{q}q}$ represents the contributions from the thermal quarks, which is
\begin{eqnarray}
\Omega _{q\bar{q}}\left( \sigma ;T,\mu \right) =-d_{q}\int \frac{d^{3}p}{%
\left( 2\pi \right) ^{3}}\{E+T\ln [ 1+e^{-\left( E-\mu \right) /T}%
]  \notag \\
 \  \ +T\ln [ 1+e^{-\left( E+\mu \right) /T}] \}.
\end{eqnarray}%
$d_{q}=12$ is the degeneracy factor of the quarks. The energy dispersion of the valence quark is $E=\sqrt{%
p^{2}+m_{q}^{2}(\sigma)}$ with the dynamical quark mass $m_{q}\left( \sigma
\right) = m_0+g\sigma$~\cite{Stephanov:2011pb,Jiang:2015hri}.  For $g=3.3$, the mass of the constituent quark is approximately
$310$ MeV, and the corresponding proton mass $m_p \sim 930 $ MeV.

\begin{figure}[tbp]
\center
\includegraphics[width=2.7 in]{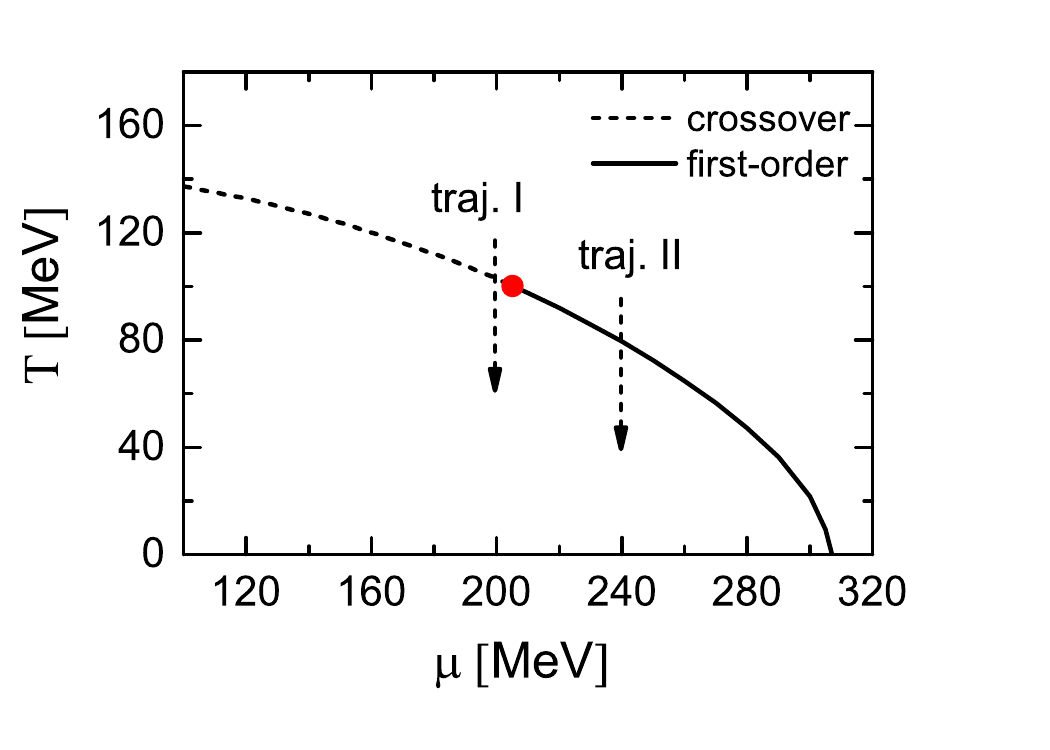}
\caption{Cartoon QCD phase diagram on the ($T, \mu$) plane.}
\label{phasediagram}
\end{figure}

According to the effective potential of Eq.\,(\ref{effpot}), the phase diagram is plotted in Fig.\,\ref{phasediagram} as the function of ($\mu, T$). The dot line denotes crossover at small $\mu$, and the solid line represents the first-order phase transition at large $\mu$. The critical point locates at $( \mu _{cp},T_{cp}) \sim \left( 205,100.2\right) $ MeV.

In Eq.\,(\ref{langevin}), the damping coefficient $\eta $ and the white noise $\xi \left( t,x\right) $ originate from the interaction between the $\sigma$ field and the heat bath, which is consisted of the thermal quarks~\cite{Biro:1997va}. In this work, $\eta$ is treated as a free parameter, within the range of models allowing, whose values $\eta=1, 3, 7$ fm$^{-1}$ are taken in the following calculations.
In the zero-momentum mode limit, the correlation of the noise has the form~\cite{Nahrgang:2011mg}
\begin{equation}
 \left\langle \xi \left( t\right) \xi \left( t^{\prime }\right) \right\rangle
= \frac{1}{V} m_{\sigma} \eta \coth{\left(\frac{m_{\sigma}}{2T}\right)} \delta \left( t-t^{\prime }\right).
\end{equation}
We remark that the zero-momentum approximation only suits the critical point scenario at the thermodynamic limit. In the realistic case with finite correlation length, we set the spatial noise at different time steps as,
\begin{eqnarray}
\xi \left( x\right) = \sqrt{\frac{1}{V} m_{\sigma} \eta \coth{\left(\frac{m_{\sigma}}{2T}\right)} \frac{1}{dt}} G\left( x\right),
\end{eqnarray}%
where $G\left( x\right)$ is a random number generator of the standard normal distribution.

For our numerical implementation, we first construct the initial profiles of the $\sigma$ field using the probability function: $  P\left[ \sigma \right] \sim \exp \left( -E \left( \sigma
\right) /T\right) $, with  $ E \left( \sigma \right) =\int d^{3}x\,
\frac{1}{2}\left( \nabla \sigma \left( x\right) \right)
^{2}+V_{\mathrm{eff}}\left( \sigma \left( x\right) \right)$.
In order to solve Eq.\,(\ref{langevin}), the space-time information of the local temperature, $T (t,x,y,z)$, and baryon chemical potential, $\mu(t,x,y,z)$, have to be known which in principle shall be extracted from the heat bath.
For simplicity, we assume the system evolves along the constant baryon density trajectories (seen traj.\,I and traj.\,II in Fig.\,\ref{phasediagram}), while the spatial-uniform temperature decreases in a Hubble-like way~\cite{Mukherjee:2015swa}:
\begin{equation}
\frac{T\left( t\right) }{T_{0}}=\left( \frac{t}{t_{0}}\right) ^{-0.45},
\end{equation}%
where $T_{0}$ is the initial temperature, and $t_{0}=1$fm is the initial time.
The whole simulation is run in a $V=6.8^3$ fm$^3$ box. The space step size $dx = dy = dz = 0.2$ fm and the time step size is $dt=0.1$ fm/c. Note that the system volume will affect the configurations of $\sigma$ field in each event, due to the change in the magnitude of the noise term, but has no influence on the results of the event-averaged $\sigma$.

With all the ingredients in hand, we complete the event-by-event simulations of the $\sigma$ field from Eq.\,(\ref{langevin}).
In the numerical calculation process, the configurations of the $\sigma$ field at every time step over $10^5$ events are recorded. The moments of the $\sigma$ field are then calculated by:
\begin{equation}
\mu _{n}^{\prime }=\langle \sigma ^{n}\rangle =\frac{\int d\sigma \sigma
^{n}P\left[ \sigma(\mathbf{x}) \right] }{\int d\sigma P\left[ \sigma(\mathbf{x}) \right] },
\end{equation}%
where $\sigma =\frac{1}{V}\int d^{3}x\,\sigma \left( \mathbf{x}\right) $.
The non-equilibrium cumulants of the $\sigma$ field are iteratively determined by the following formula:%
\begin{eqnarray}
C_{1} &=&\mu _{1}^{\prime },  \label{eqn_def_C1}\\
C_{2} &=&\mu _{2}^{\prime }-\mu _{1}^{\prime 2},   \label{eqn_def_C2}\\
C_{3} &=&\mu _{3}^{\prime }-3\mu _{2}^{\prime }\mu _{1}^{\prime }+2\mu
_{1}^{\prime 3},   \label{eqn_def_C3}\\
C_{4} &=&\mu _{4}^{\prime }-4\mu _{3}^{\prime }\mu _{1}^{\prime }-3\mu
_{2}^{\prime 2}+12\mu _{2}^{\prime }\mu _{1}^{\prime 2}-6\mu _{1}^{\prime 4}.\label{eqn_def_C4}
\end{eqnarray}

\section{Numerical results and discussions}\label{sec:result}
\subsection{dynamical evolution in the crossover scenario
--- critical slowing down}

\begin{figure*}[t]
\center
\includegraphics[width=2.0 in, height=1.6 in]{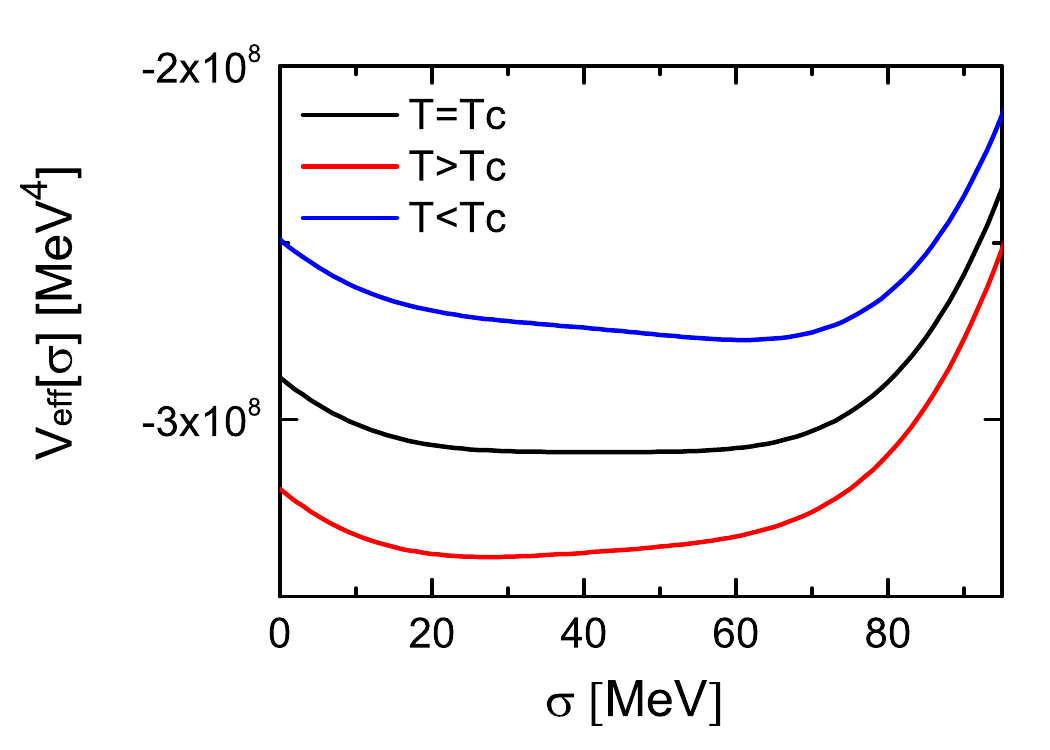}
~~~~~~~\includegraphics[width=2.0 in, height=1.6 in]{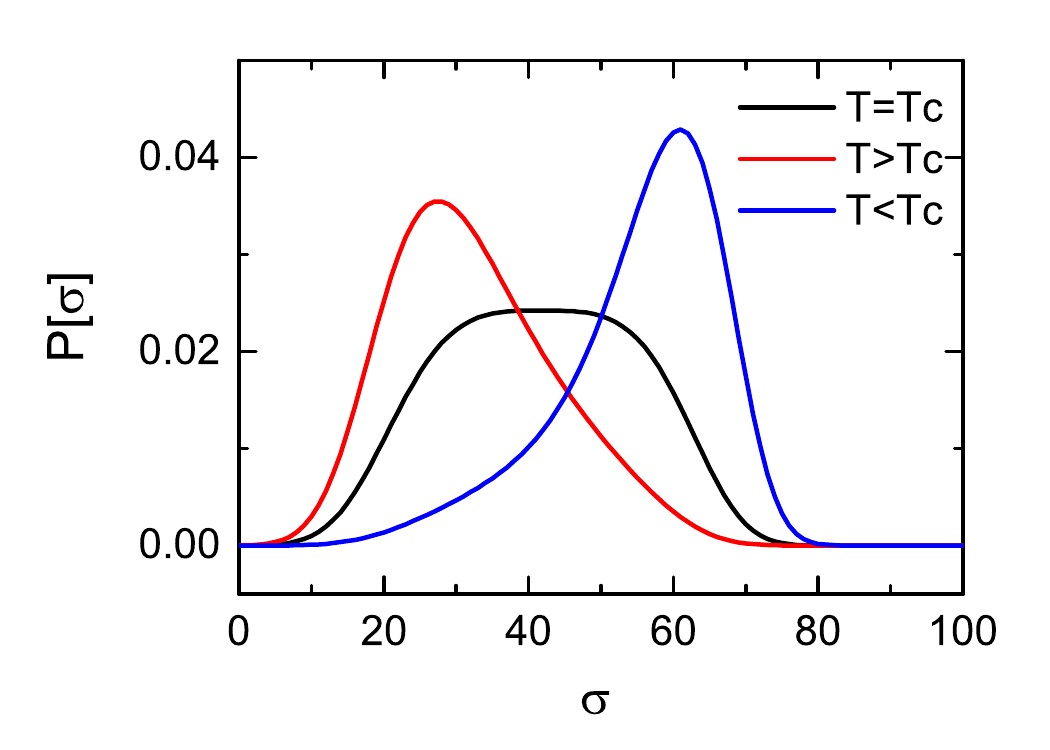}
\caption{Given $\mu = 200 $ MeV, the effective potential and probability distribution as functions of $\sigma$ at different temperature.}
\label{mu=205}
\end{figure*}

\begin{figure*}[t]
\center
\includegraphics[width=3.2 in]{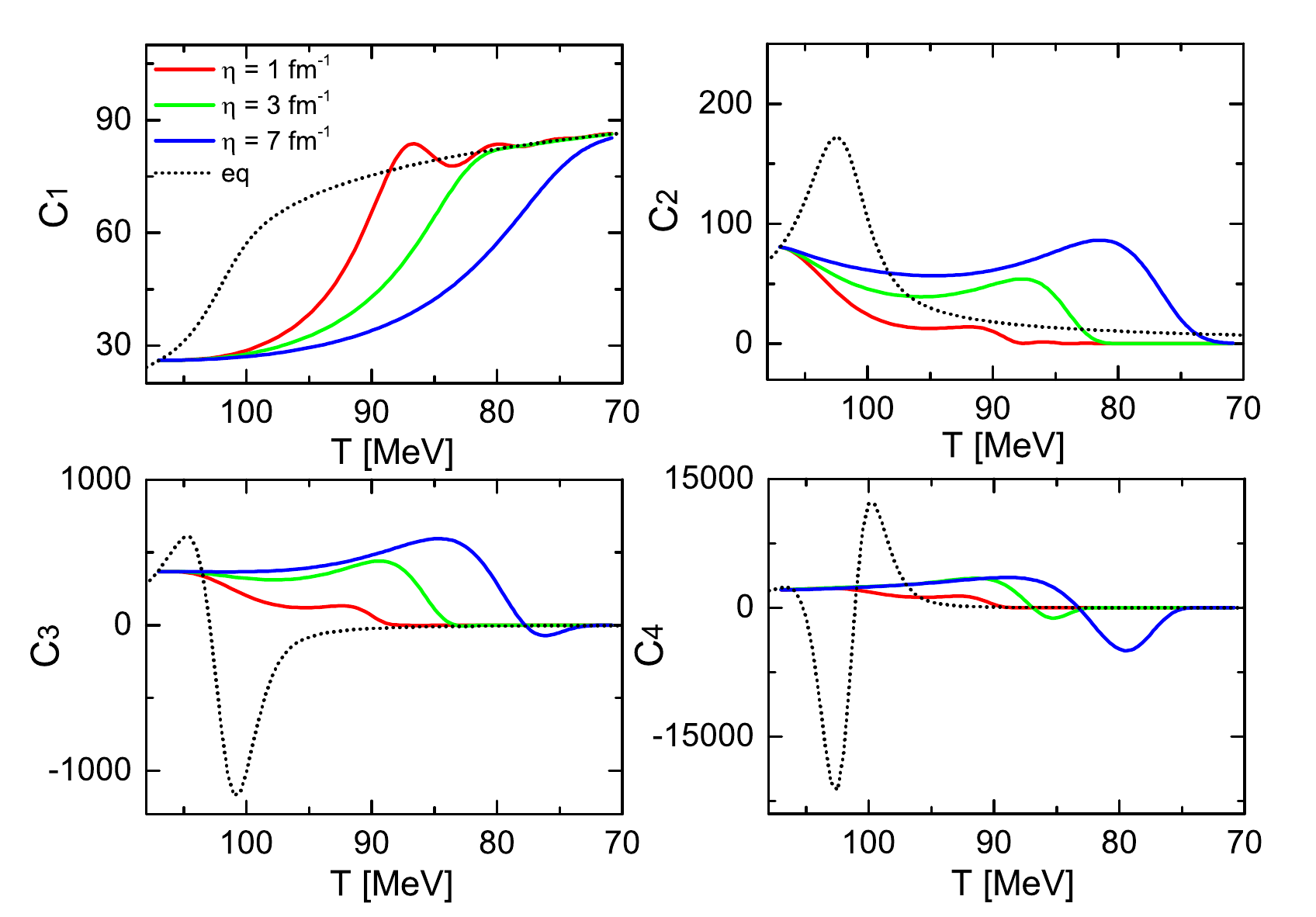}
\includegraphics[width=3.2 in]{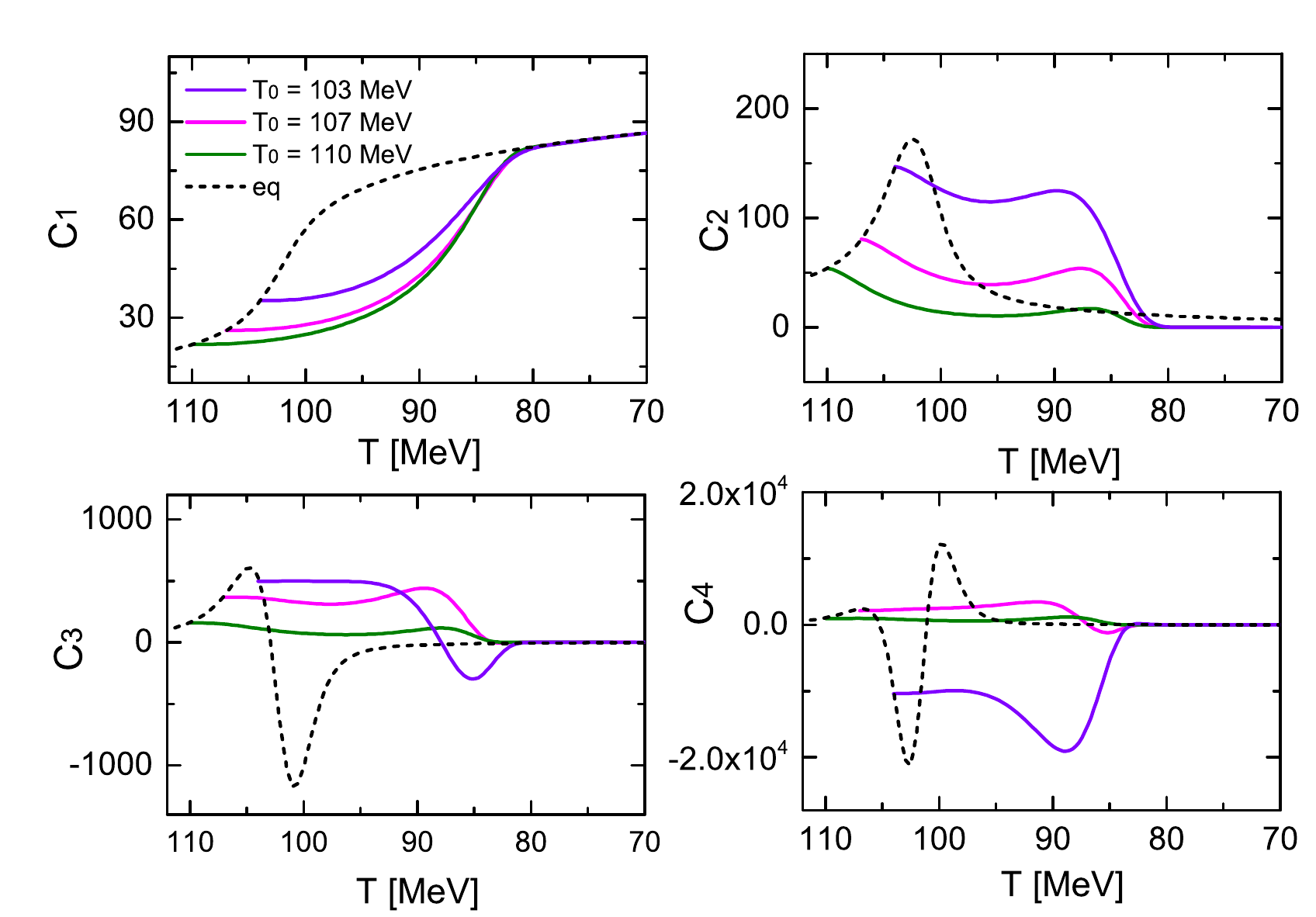}
\caption{Non-equilibrium cumulants of $\sigma$ field (units of $C_n$: MeV$^n$.) along traj.\,I. The dashed lines denote the equilibrium values, and the colored lines represent the non-equilibrium cumulant values. Left panel: Cumulant evolution of $\sigma$, which starts at the same temperature but develops under different damping coefficients. Right panel: given $\eta =3\,\mathrm{fm}^{-1}$, cumulative evolution of $\sigma$s at different initial temperature (initial configurations).}
\label{IPO=2}
\end{figure*}

We first solve the Langevin equation in the crossover scenario for $\mu = 200$ MeV. The effective potential and the probability distribution function for the phase transition region are shown in Fig.\,\ref{mu=205}.
Both the effective potential and the distribution function each have a dip or peak at the given temperature region. The system smoothly transits from the symmetry restored phase to the symmetry broken phase as the temperature decreases.

In Fig.\,\ref{IPO=2},
we plot the dynamical evolution of $\sigma$'s cumulants along traj.\,I as marked on the cartoon phase diagram of Fig.\,\ref{phasediagram}. The corresponding horizontal axis is the inverted temperature as time increases.
The dashed lines represent the equilibrium cumulants, and the colored solid lines represent the non-equilibrium ones at a given damping coefficient. In the left panel, we show the behavior of $\sigma$'s cumulants at different damping coefficients.
The initial configurations of $\sigma$ fields are constructed to satisfy the equilibrium distribution at $t_{0}$, thus both the non-equilibrium and equilibrium cumulants have the same values at the starting point.
In the evolution, the non-equilibrium cumulants present clear memory effects, going after the trends of equilibrium ones as temperature decreases and reaching their maxima (or minima) at later times.
Finally, after the phase transition, the effective potential changes from a non-Gaussian shape to a Gaussian shape.
As expected, the non-equilibrium $C_{1}$ goes to the equilibrium value and the high-order cumulants vanish at the broken phase.
During the expansion, non-equilibrium $C_2$ decreases slightly at the earlier stage and then grows due to the broadening of the effective potential in the critical region. We emphasize that the signs and values of the dynamical $C_{3,4}$ strongly differ from the equilibrium ones in a large T region below $T_c$.
For larger damping coefficients, the $\sigma$  field relaxes as slowly as it should and the system takes a longer time to approach equilibrium. Prolonging the duration to its out-of-equilibrium state, the maximum (or minimum) of high-order cumulants are enhanced as $\eta$ increases.
The effect of the damping coefficients can be estimated in the strong limit. In the critical region, the behavior of the critical mode simplifies to a diffusion-like process, as $\omega\ll\eta$, the higher order time derivative term are ignored. Shown in Fig.\ref{diffusion} in Appendix 1, a large $\eta$ results in an over-damped system, and the cumulants from the Langevin dynamics match those from the diffusion equation.

Besides the damping coefficients, the initial conditions also have a significant influence on the magnitude of the dynamical cumulants. In the right panel of Fig.\,\ref{IPO=2}, we exhibit the results of non-equilibrium cumulants starting from three different initial temperatures above $T_c$, with the damping coefficient fixed at $\eta = 3~\mathrm{fm}^{-1}$.
The initial $\sigma$ field configurations are again sampled according to the equilibrium distribution function and thus vary for different $T_0$.
Again, the initial values of the $\sigma$'s cumulants are governed by the $\sigma$ field's distribution at starting temperature. We find that the high-order cumulants are strongly enlarged while $T_0$ are close to $T_{c}$ and maintain substantial values during the later non-equilibrium development until below $T_c$.
In addition, with the same damping coefficient, the cumulants reach their maxima (or minima) at approximately the same temperature, which is almost unrelated to their starting points.

Note that we have also carried out calculations on the dynamical cumulants both with and without the spatial fluctuations of the $\sigma$ fields, as demonstrated in Appendix 2. The influence of spatial fluctuations is diminished since we perform the statistics based on the event-averaged $\sigma$ field. Shown in Fig.\,\ref{flucs}, for the specified values of $\mu$ and $T$, we compare $\sigma$'s $C_i$ at two cases. As evident, it is challenging to observe the impact of the spatially non-uniform $\sigma$ field on the cumulants after event-by-event averaging.

\subsection{dynamical evolution in the first-order phase transition scenario
 --- supercooling effect}
\begin{figure*}
\center
\includegraphics[width=2.0 in, height=1.6 in]{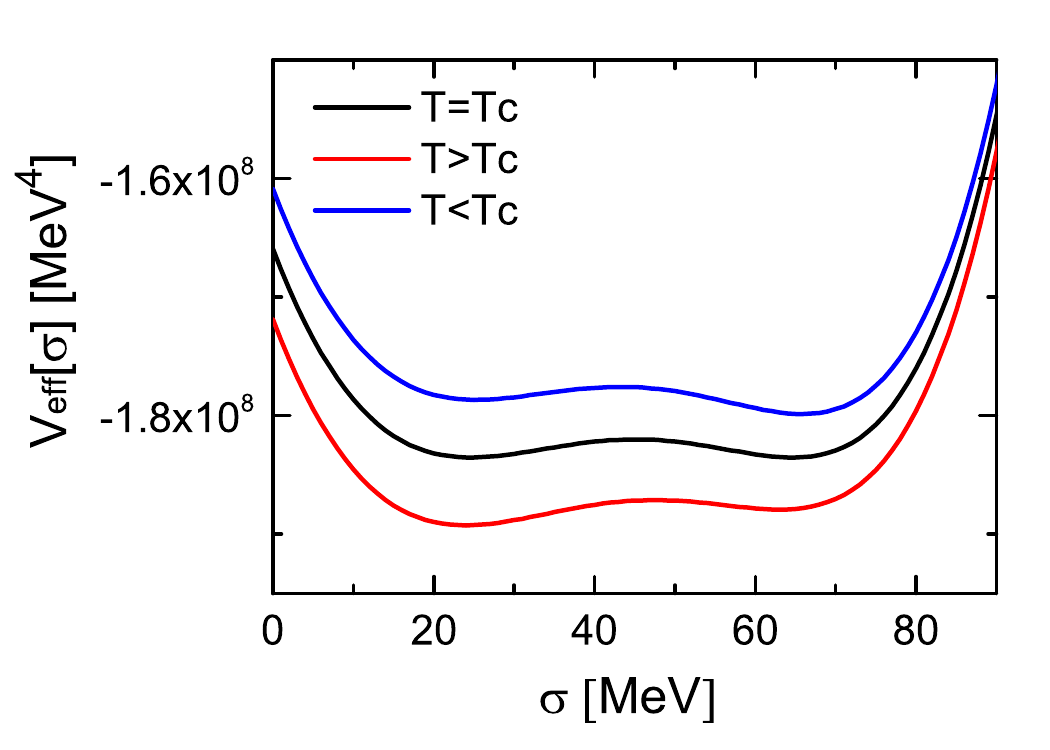}
~~~~~~~~~\includegraphics[width=2.0 in, height=1.6 in]{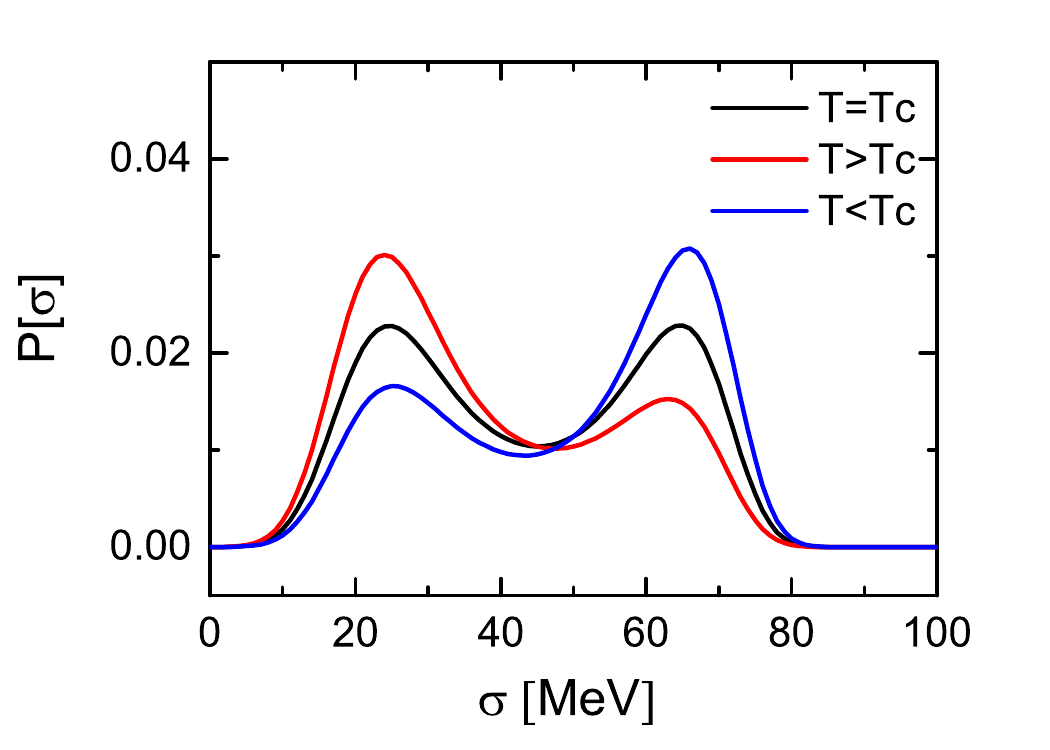}
\caption{Given $\mu = 240 $ MeV, the effective potential and probability distribution as functions of $\sigma$ at different temperatures.}
\label{mu=240}
\end{figure*}

\begin{figure*}
\center
\includegraphics[width=3.2 in]{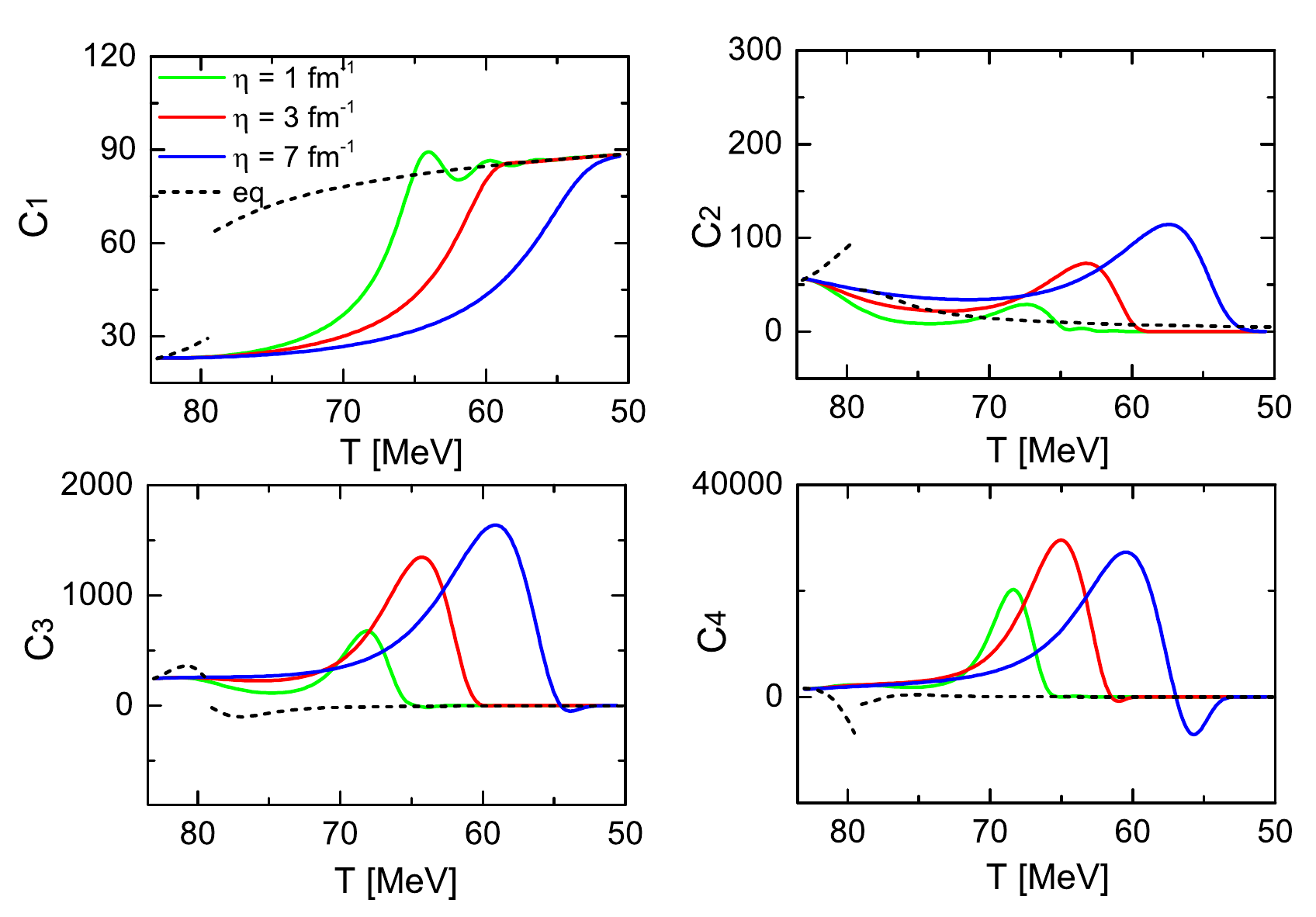}
\includegraphics[width=3.2 in]{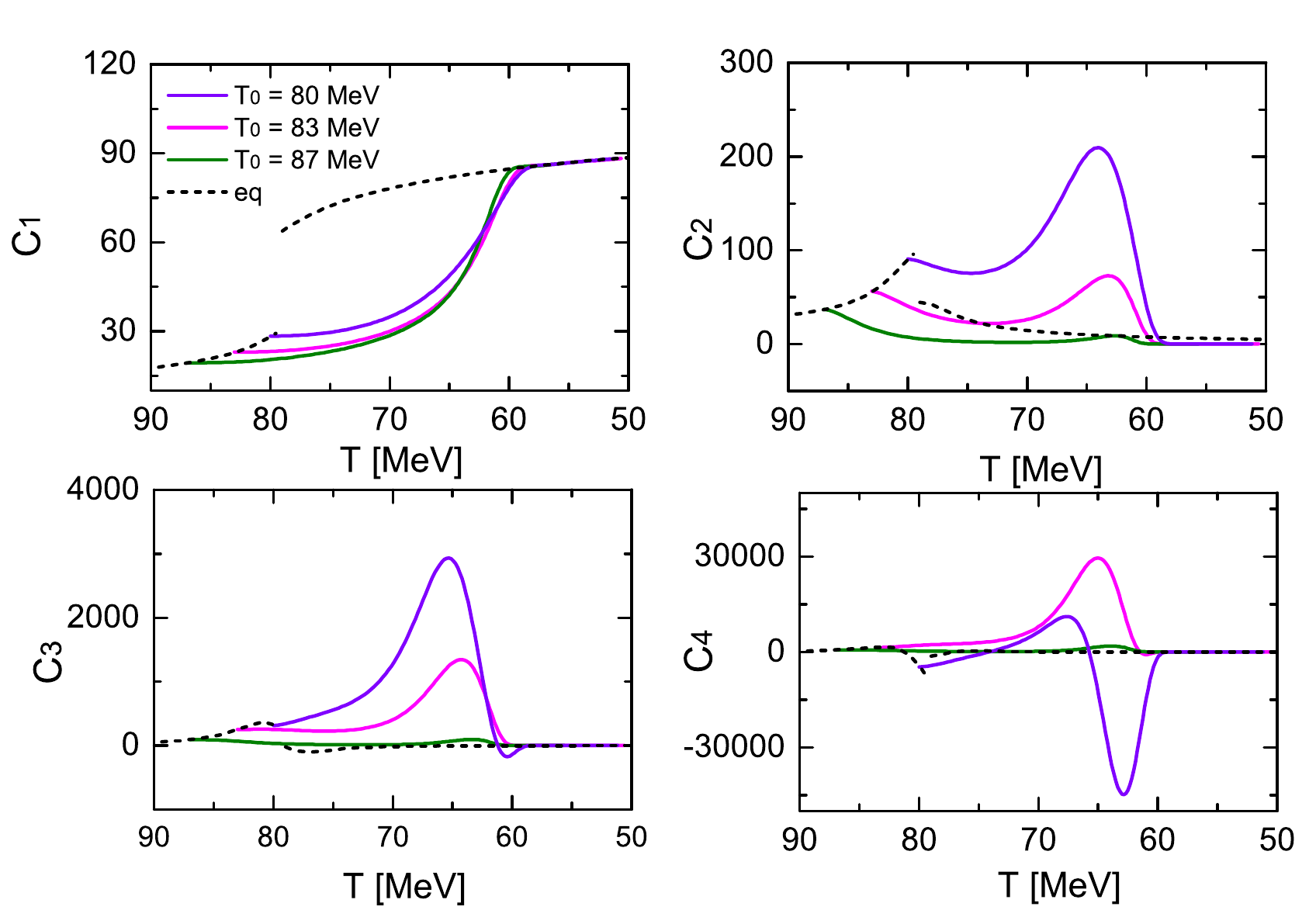}
\caption{Non-equilibrium cumulants of $\sigma$ field along traj.\,II.
Left panel: Cumulant evolution of $\sigma$, starting at the same temperature but developing under different damping coefficients. Right panel: given $\eta = 3\,\mathrm{fm}^{-1}$, the cumulative evolution of $\sigma$'s at different initial temperature (initial configurations).}
\label{IPO=1}
\end{figure*}

In this subsection, we discuss the dynamical evolution of $\sigma$'s cumulants in the first-order phase transition scenario, with the baryon chemical potential fixed at $\mu =240 MeV$.
As shown in the left panel of Fig.\,\ref{mu=240}, the thermodynamic potentials are characterized by two co-existing phases near $T_c$.
At the thermodynamic limit, the probability distribution function is double $\delta$-like only at the phase transition point. The $\sigma$ field stays at the global minimum of the effective potential, and there is a discontinuity at the phase transition temperature in the equilibrium cumulants. In turn, confining to a finite system volume ($V=6.8^3$ fm$^3$), the probability distribution function presents two peaks with comparable probability in the phase transition region (as shown in the right panel of Fig.\,\ref{mu=240}), which leads to significantly different behaviors of the high-order cumulants.

In Fig.\,\ref{IPO=1}, we present the numerical results of dynamical cumulants as functions of decreasing temperature, as denoted by the traj.\,II  in Fig.\,\ref{phasediagram}. The left panel represents the evolution of cumulants starting from a given set of $\sigma$'s configurations under different damping coefficients. For fixed $\eta = 3~\mathrm{fm}^{-1}$, the right panel shows the cumulative behaviors starting at various initial temperatures. Similar to the case in the crossover scenario, the diffusive dynamics render the same memory effects for the non-equilibrium cumulants. In addition, the non-equilibrium cumulants are continuous and much larger than that of the equilibrium ones.

The significant enhancements of the non-equilibrium cumulants are explained in the following.
In the first-order phase transition scenario, the existence of a barrier between the two minima in the thermodynamic potential prevents the $\sigma$'s configurations from shrinking to the global minimum even when the temperature is lower than $T_c$ (known as supercooling effects in thermodynamics). Since the $\sigma$ field is trapped in the original minima during the cooling down process, only the events with intense thermal fluctuations would overcome the potential barrier.
Then as the broadening of the probability distribution function in a finite-size system, different events occupy both of the states with comparable weights, which leads to a strong departure of $C_{3,4}$ from the equilibrium cumulants. Such enhancements of cumulants at the boundary side of the first-order phase transition have the potential to address the large deviations of BES data $\kappa \sigma^2$ from the statistical baselines at low collision energies.

\subsection{cumulants on the hypothetical freeze-out lines}

The possible signals of phase transition are measured after the particles chemically freeze out, and in this subsection, we discuss the non-equilibrium cumulants' behaviors on the hypothetical freeze-out lines as functions of baryon chemical potential.
Within the present model setting, the phase transition line described by the linear sigma model is far away from the freeze-out line determined by the statistical model via fitting experimental data~\cite{Andronic:2017pug}.
Thus, we are not able to directly borrow the freeze-out information from the experiments. In the following, we artificially choose the freeze-out lines and assume the dynamical evolution of the $\sigma$ field along different trajectories starting at $T_0 = (T_c + 4 )$~MeV, and freezing out at either $T_{\mathrm{f1}}= (T_0 - 10) $~MeV or $T_{\mathrm{f2}} = (T_0 - 15) $~MeV. In the duration of each trajectory, the baryon chemical potential is fixed.

\begin{figure*}[tbp]
\center
\includegraphics[width=3.3 in]{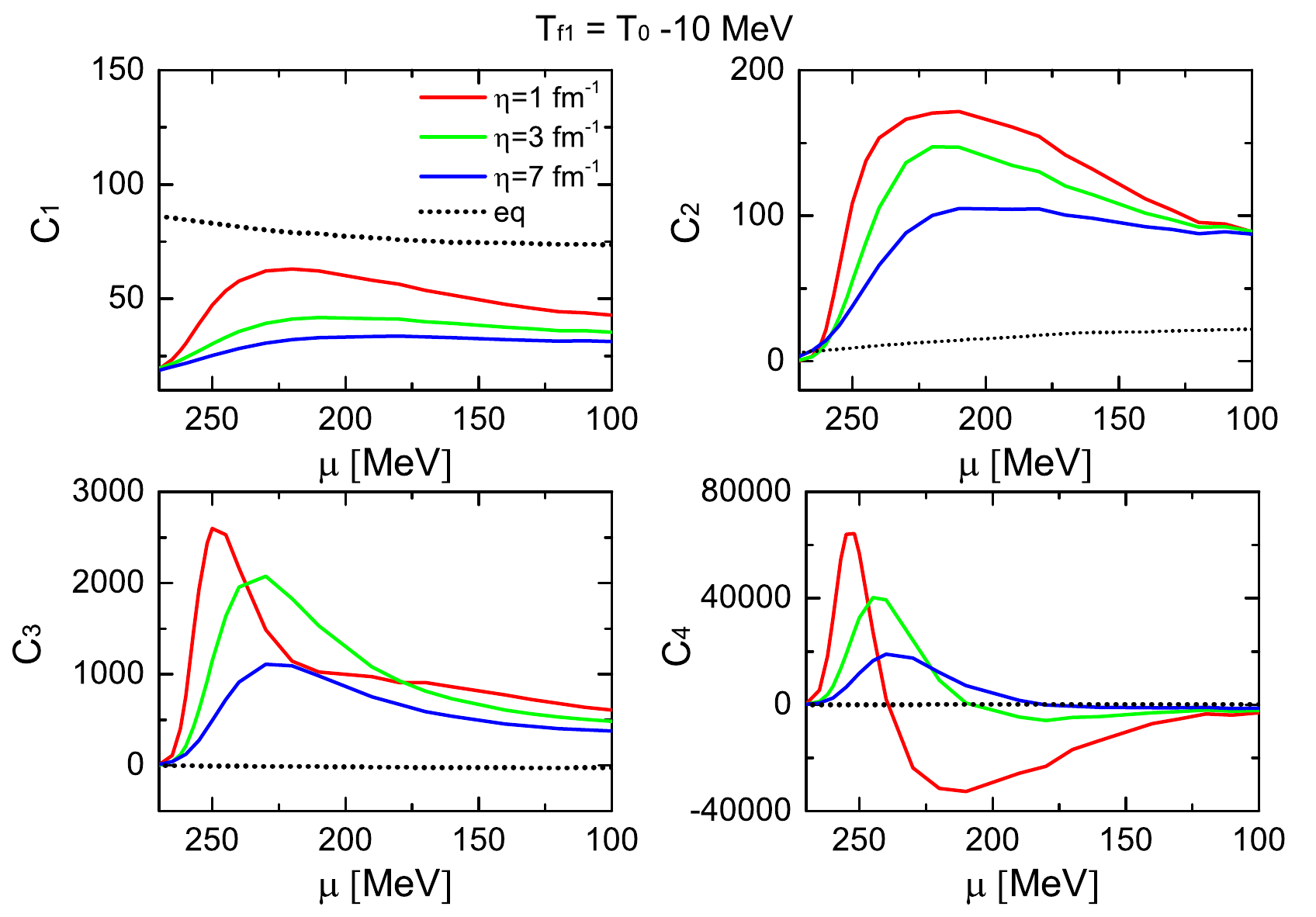}
\includegraphics[width=3.3 in]{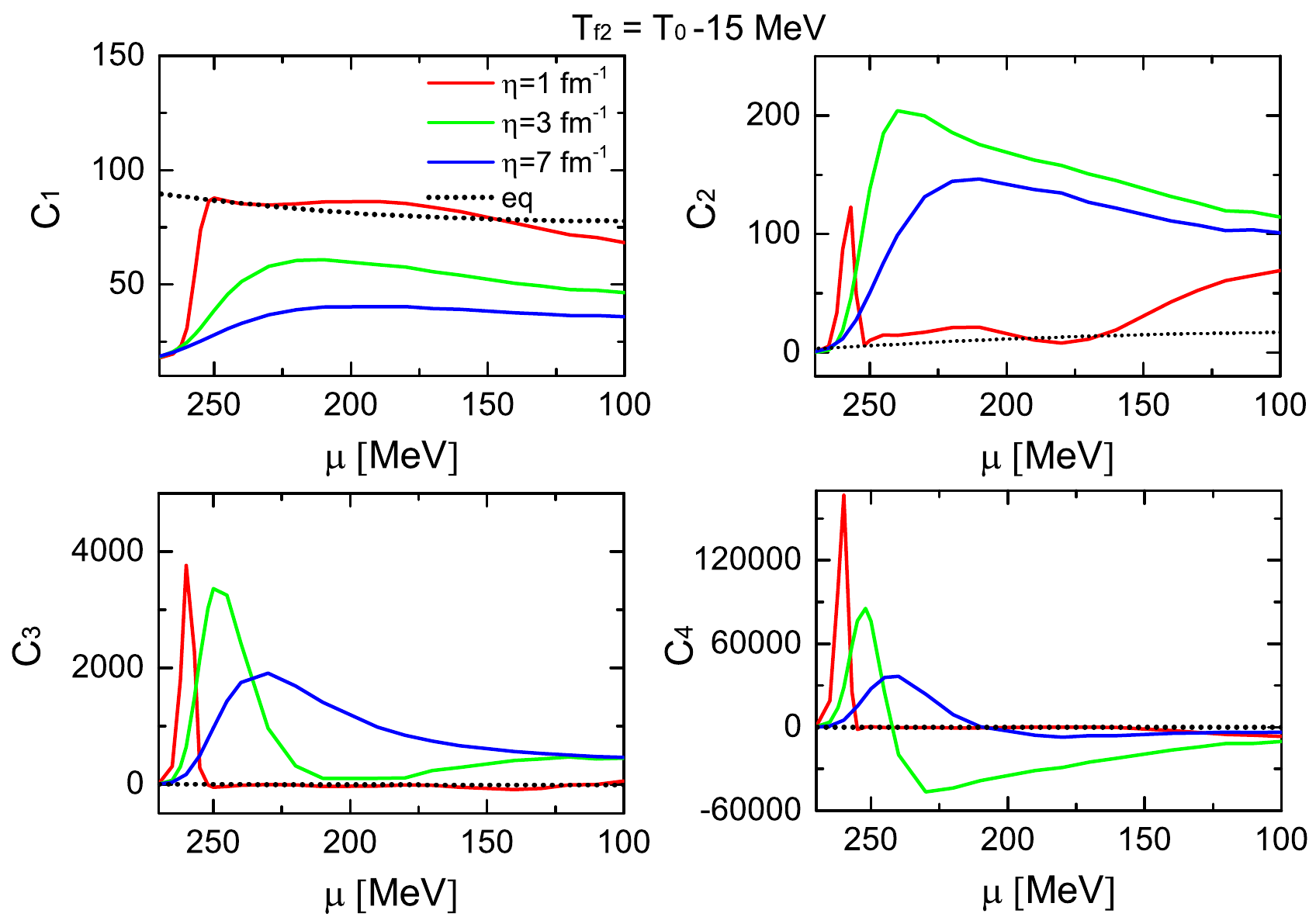}
\caption{$\sigma$'s cumulants as functions of baryon chemical potentials, ending on the hypothetical freeze-out line with $T_{\mathrm{f1}}
= (T_0 - 10 )$~MeV (left panel) and $T_{\mathrm{f2}} = (T_0 - 15) $~MeV (right panel).}
\label{Tf-sigma-10}
\end{figure*}

In Fig.\,\ref{Tf-sigma-10}, we draw the $\sigma$'s cumulants as functions of the baryon chemical potential, adopting  $T_{\mathrm{f1}}$ and $T_{\mathrm{f2}}$ as the freeze-out temperature individually.
The high-order equilibrium cumulants decay to zero as the system is away from the phase transition region. Rather, the non-equilibrium cumulants exhibit large deviations from the equilibrium ones and significant non-monotonic structures on the hypothetical freeze-out line.
Below $T_{c}$, the equilibrium $C_3$ is negative and limited to zero as temperature decreases. However, the non-equilibrium $C_3$ is positive in most phase space of $\mu$ for both applied freeze-out lines. The flipping of signs could address the sign problem based on the prediction of equilibrium critical fluctuations~\cite{Jiang:2015hri}.
Last but not least, the non-equilibrium $C_4$ is oscillating near the critical baryon chemical potential ($\sim 205~$MeV) and tends to zero around $\mu \sim 270~$MeV.

By comparing the vanishing equilibrium cumulants of $C_{3,4}$, the dynamical processes provide us with abundant nontrivial behaviors. In our model calculation, the appearance of the non-monotonic curves of $C_{4}$ originates from a combined effect of $\eta$ and the hypothetical freeze-out line. At the first freeze-out temperature $T_{\mathrm{f1}}$, the evolution of $C_{4}$ at smaller $\eta$ (red and green lines) show the non-monotonic structure, but they oscillate under the influence of larger $\eta$ (green and blue lines) at freeze-out temperature $T_{\mathrm{f2}}$. This means that $C_{4}$ evolves following its own dynamical processes and its behavior is non-universal.
Furthermore, the deviation of $C_{4}$ not only comes from the development of itself but also from the non-equilibrium features of other cumulants since the higher order cumulants are coupled to the lower order ones as shown in Eq.(\ref{eqn_def_C1}-\ref{eqn_def_C4}).

The peaks of $C_3$ and $C_4$ are worth paying attention to, as well. Both of these maximums take place at a critical value around $\mu\sim 240~$ MeV. Such maximums are induced by the maximization of the supercooling effect in the current model.
With $\mu$ slightly larger than $\mu_c$, the barrier between the global minima and the false minima prevents a critical number of events from developing to the global minimum. The $\sigma$ field is approximately evenly distributed in both minima for different events, and induces a dramatically large peak of the high-order cumulants. At the high baryon chemical potential, $\mu \sim 270$ MeV, the barrier is so strong that the $\sigma $ field is trapped in the original minima and can not escape even at a temperature much lower than $T_c$.
Without the fluctuations manifesting the phase transition, the high-order cumulants are suppressed, and their magnitude approaches the equilibrium limit.

In spite of the various cumulative behaviors found in the first-order phase transition scenario, we note that one should be cautious when comparing the current model simulation at a high baryon chemical potential region to the experimental data, since a mismatch between the model and the experiments at small collision energies is likely to occur. In this work, we have presumed that system evolution occurs when the chiral symmetry is restored, but quark-gluon-plasma may not be created in heavy-ion collisions at low collision energies. A sophisticated first-order phase transition with the proper degree of freedom should be further explored and studied in future work.

\section{Summary and outlook}\label{sec:con}

To summarize, in this work, we study the dynamical evolution of the $\sigma$ field based on the event-by-event simulations of a single component Langevin equation.
The temperature decrease for the system is set to be Hubble like and the computation is completed in a finite-size system. We statistically weight the dynamical variable $\sigma$ over $10^5$ events during the real-time evolution to obtain its high-order cumulants.
With the current model setting, we find that the non-equilibrium cumulants express clear memory effects, and the magnitude of $C_2$ slowly increases as the system approaches its critical regions. The signs of $C_{3,4}$, as well as the magnitudes, can differ from the equilibrium ones below $T_{c}$. We also find that the high-order cumulants are significantly enhanced on the boundary side of the first-order phase transition.
Finally, the spread of out-of-equilibrium cumulants along the hypothetical freeze-out lines has been presented. The non-equilibrium $C_3$ is positive at large baryon chemical potentials, in contrast to the negative sign of the equilibrium $C_{3}$. In the vicinity of CP, with certain parameter sets, the non-equilibrium $C_4$ expresses non-monotonic curves at large $\mu$ region. We conclude that the combination of supercooling effect and dynamical effects on the first-order phase transition side plays a dominant role in the nonmonotonicity of the high-order cumulants on the hypothetical freeze-out lines.

Note that the apparent memory effects of the cumulants up to the fourth order based on the Langevin framework are in accord with those obtained by the use of the Fokker-Plank equations \cite{Mukherjee:2015swa,Jiang:2017sni}, but currently not sufficiently discussed and presented elsewhere, which is crucial for the study of nonequilibrium signals from experiments. Of course, in order to quantitatively describe the experimentally observed cumulants of net proton multiplicities, more sophisticated and realistic dynamical modeling are required from the theoretical side~\cite{Kapusta:2011gt,Akamatsu:2017rdu}. A suitable mathematical tool that can accurately describe the properties of fluids in heavy-ion collisions and better performs numerical calculations is still under exploration. It is well known that the method of Fokker-Plank equations is equivalent to the Langevin dynamics in the Markov process~\cite{Mukherjee:2015swa,Jiang:2017sni,An:2021wof}. However, taking into account the realistic fireball evolution in heavy-ion collisions, where the spatial distributions of temperature and baryon chemical potential, etc., are nonuniform, the Langevin dynamics has the advantage of easily including those effects in simulations, and further combines with the hydrodynamic equations to describe the complete dynamical process in RHIC. We note here how these non-Markov effects manifest in the Fokker-Plank equations is beyond the scope of this paper. Further investigation and analysis are under work.

In this paper, with a relatively simple setup for the modeling and the parameters, we present the dynamics of  the cumulants, which serves as reference information for the analysis of the experimental measurements. For the explanation of the experimental data, a number of effects are playing their roles, including the subject of the proper equation of state~\cite{Monnai:2019hkn,Noronha-Hostler:2019ayj,Parotto:2018pwx}, of the unknown parameters of the Ising-to-QCD mapping~\cite{Mroczek:2020rpm}, of the critical transport coefficients~\cite{Karpenko:2015xea,Denicol:2015nhu,Bernhard:2019bmu,Martinez:2019bsn}, of the finite size, finite size scaling and global charge conservation in the vicinity of a CP~\cite{Sombun:2017bxi,Antoniou:2017vti,Vovchenko:2020gne,Poberezhnyuk:2020ayn,Tomasik:2021jfd}, of the non-critical baselines for the cumulants of net-proton number fluctuations~\cite{Braun-Munzinger:2020jbk}, of the nonuniform temperature/chemical potential effects~\cite{Zheng:2021pia}, and of the proper freezeout scheme in the critical region~\cite{Pradeep:2022mkf}.
Besides, further connections between the criticality and the other experimental observable are being established through  theoretical efforts~\cite{Mukherjee:2016kyu,Sun:2017xrx,Sun:2018jhg,Wu:2018twy,Brewer:2018abr,Pratt:2019fbj} and new technique such as the machine learning method~\cite{Pang:2016vdc,Huang:2018fzn,Jiang:2021gsw} is also being developed for the search of QCD phase transition signals as well.

\section*{APPENDIX}
\subsection*{1. the comparison of dynamical results from diffusion equation and
Langevin equation}\label{appendix1}

\begin{figure}[t]
\center
\includegraphics[width=3.5 in]{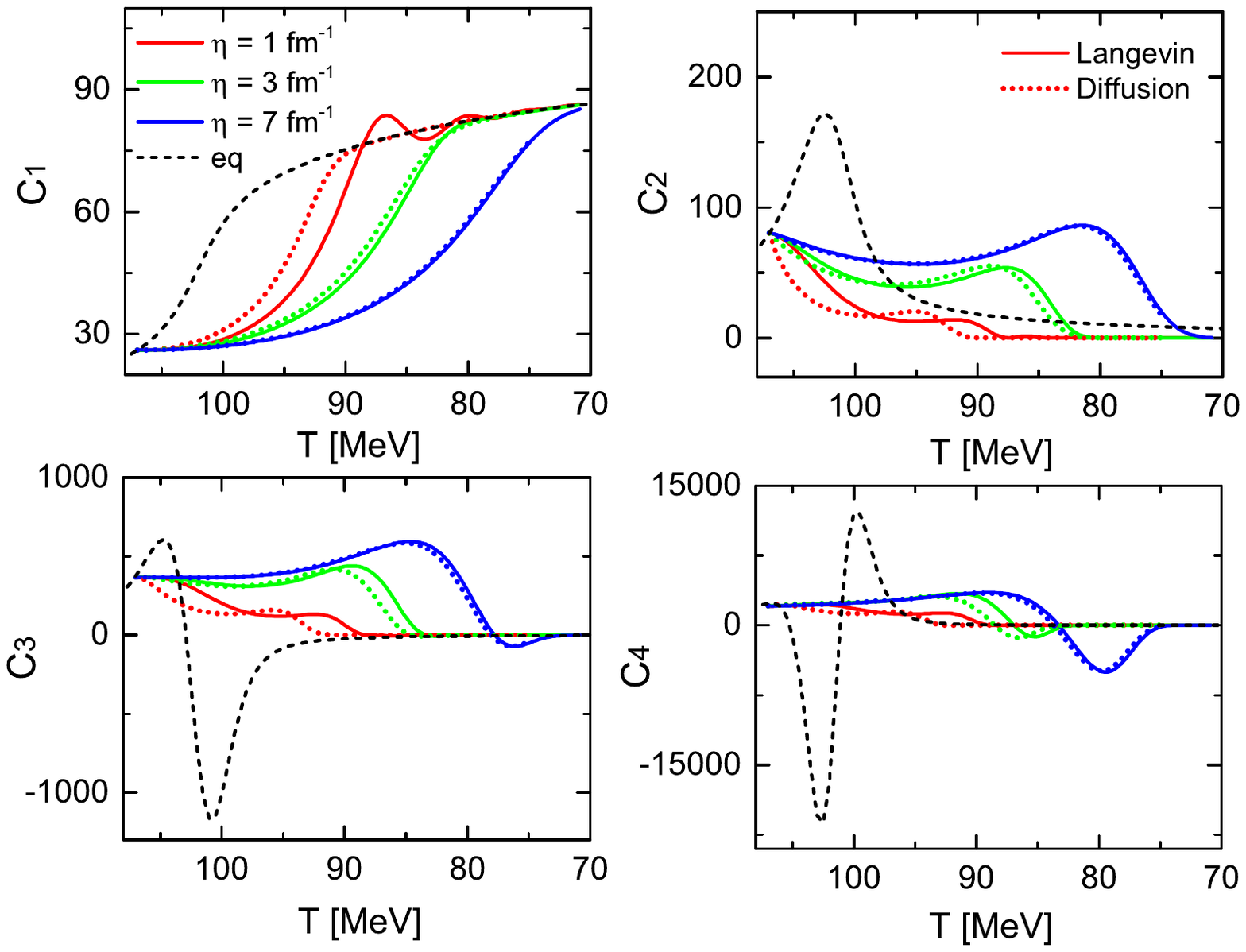}
\caption{$\sigma$'s cumulants as functions of temperature, rendered by either the Langevin equation (solid lines) or the over-damped diffusion equation (dotted lines).}
\label{diffusion}
\end{figure}
In the low frequency range, the evolution of the sigma field is described by an over-damped diffusion-like equation, since $\omega\ll\eta$. Ignoring the second order time derivative term, the diffusion equation takes the form:
\begin{equation}
\eta \partial_{t}\sigma \left( t,x\right) - \nabla ^{2}\sigma \left(
t,x\right) +\frac{\delta V_{eff}\left( \sigma \right) }{\delta \sigma }=\xi
\left( t,x\right).
\end{equation}
Here we compare the numerical evolution of the sigma field using the Langevin equation as well as the over-damped diffusion equation.
In Fig.\,\ref{diffusion}, we plot the Langevin (solid lines) dynamics of $\sigma$'s cumulants and the over-damped diffusion equation (dotted lines) with three different damping coefficients at  $\mu = 200$ MeV. In the figure, it is shown that the differences of $\sigma$'s cumulants decrease with an increase in the damping coefficient. At $\eta = 7~\mathrm{fm}^{-1}$, the evolution differences between the two kinds of equations can be safely ignored.

\subsection*{2. the comparison of dynamical cumulants with and without spatial fluctuations}\label{appendix2}

\begin{figure}[t]
\center
\includegraphics[width=3.5 in]{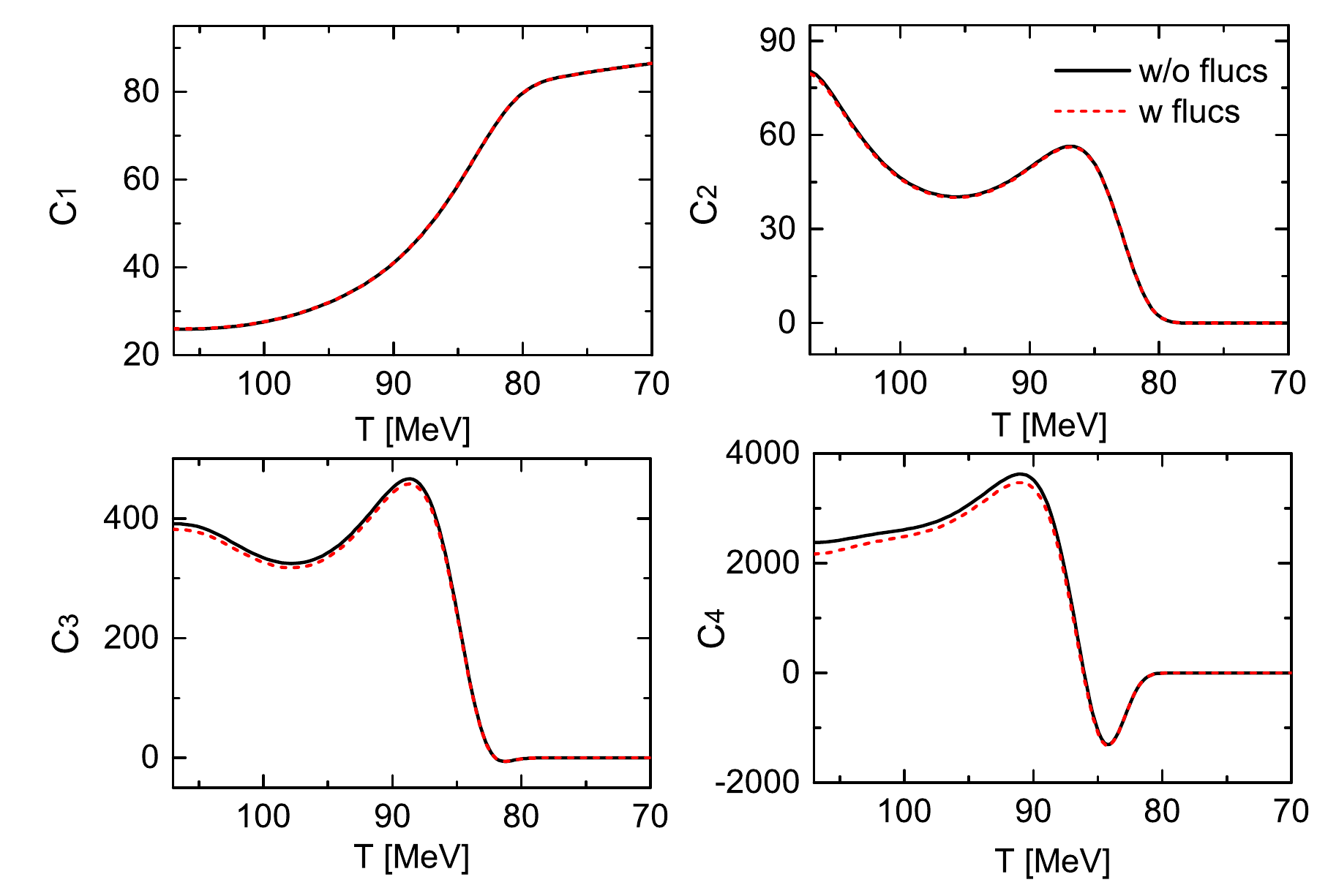}
\caption{Given $\mu=200 $ MeV and $\eta = 3\mathrm{fm}^{-1}$, the black lines represent cumulants calculated from sigma’s configurations without fluctuations, and the red dashed lines represent results with fluctuations.}
\label{flucs}
\end{figure}

Here, we numerically simulate the dynamical cumulants under the assumption that the sigma field is spatially homogeneous  (the black lines in Fig.\,\ref{flucs}) and compare the results to the cases with spatial fluctuations (red dashed lines in Fig.\,\ref{flucs}).
It has been discovered that there are no apparent differences between the first and second-order cumulants. The third and fourth-order cumulants have larger magnitudes in the early stages for the spatially uniform $\sigma$ events, but the differences reduce in later stages due to the damping of the $\sigma$ field. Thus based on the above calculations, we find that it is difficult to detect the impact of $\sigma$'s spatial fluctuations on the event-averaged quantities.

\section*{Acknowledgements}

We thank Ulrich. Heinz, Yu-Xin Liu, Swagato Mukherjee, Misha Stephanov, Derek Teaney,  Yi Yin, Shanjin Wu and Huichao Song for useful discussion and comments. We thank the anonymous reviewer for his/her valuable suggestions in the impact of spatial non-uniformity of the sigma field. The work of LJ has been supported by the NSFC under grant no. 12105223, and the work of JC has been supported by the start-up funding from Jiangxi Normal University under grant No. 12021211.

\end{document}